\documentclass[nofootinbib,twocolumn,showpacs,preprintnumbers,pre,aps,superscriptaddress]{revtex4-1}

\usepackage[dvipdfmx]{graphicx}
\usepackage{bm}
\usepackage{amsmath}
\usepackage{amssymb}
\usepackage{txfonts}
\usepackage{color}
\usepackage{tikz}
\usepackage{esint}

\DeclareMathOperator{\sech}{sech}

\begin{document}

\title{Thermally Driven Two-Sphere Microswimmer with Internal Feedback Control}

\author{Jun Li}
\affiliation{Department of Physics, Wenzhou University, Wenzhou, Zhejiang 325035, China}
\affiliation{Wenzhou Institute, University of Chinese Academy of Sciences, 
Wenzhou, Zhejiang 325001, China}

\author{Ziluo Zhang}
\affiliation{Wenzhou Institute, University of Chinese Academy of Sciences, 
Wenzhou, Zhejiang 325001, China}
\affiliation{Institute of Theoretical Physics, Chinese Academy of Sciences, 
Beijing 100190, China}

\author{Zhanglin Hou}
\affiliation{Wenzhou Institute, University of Chinese Academy of Sciences, 
Wenzhou, Zhejiang 325001, China} 
\affiliation{Oujiang Laboratory, Wenzhou, Zhejiang 325000, China}

\author{Yuto Hosaka}
\affiliation{Max Planck Institute for Dynamics and Self-Organization (MPI DS), 
Am Fa{\ss}berg 17, 37077 G\"{o}ttingen, Germany}

\author{Kento Yasuda}
\affiliation{Research Institute for Mathematical Sciences, 
Kyoto University, Kyoto 606-8502, Japan}

\author{Linli He}\email{Corresponding author: linlihe@wzu.edu.cn}
\affiliation{Department of Physics, Wenzhou University, Wenzhou, Zhejiang 325035, China}

\author{Shigeyuki Komura}\email{Corresponding author: komura@wiucas.ac.cn}
\affiliation{Wenzhou Institute, University of Chinese Academy of Sciences, 
Wenzhou, Zhejiang 325001, China} 
\affiliation{Oujiang Laboratory, Wenzhou, Zhejiang 325000, China}


\begin{abstract}
We investigate the locomotion of a thermally driven elastic two-sphere microswimmer with internal feedback 
control that is realized by the position-dependent friction coefficients.
In our model, the two spheres are in equilibrium with independent heat baths characterized by different 
temperatures, causing a heat flow between the two spheres. 
We generally show that the average velocity of the microswimmer is non-zero when the friction coefficients 
are dependent on the spring extension.
Using the method of stochastic thermodynamics, we obtain the entropy production rate and  
estimate the efficiency of the two-sphere microswimmer.
The proposed self-propulsion model highlights the importance of information in active matter and 
offers a fundamental transport mechanism in various biological systems.
\end{abstract}

\maketitle

\section{Introduction}
\label{introduction}

Microswimmers, such as sperm cells or motile bacteria, are tiny objects moving in viscous environments,  
and are expected to be relevant to microfluidics and microsystems~\cite{LaugaBook}. 
By transforming chemical energy into mechanical work, microswimmers change their shapes and move in 
viscous fluids~\cite{Hosaka22}.
According to Purcell's scallop theorem, microswimmers in a Newtonian fluid need to undergo non-reciprocal 
body motion for steady locomotion~\cite{Purcell1977,Ishimoto12}.
To realize such a non-reciprocal deformation, various microswimmer models that have more than 
two degrees of freedom have been proposed~\cite{Lauga09a}.
One example is the three-sphere microswimmer in which three in-line spheres are linked by two arms of 
varying lengths~\cite{Najafi04,Golestanian08}.

Among various generalizations of the three-sphere microswimmer model~\cite{Yasuda23}, 
Hosaka \textit{et al.} proposed an elastic three-sphere microswimmer in which the three spheres are in 
equilibrium with independent heat baths having different temperatures~\cite{Hosaka17}.
It was shown that such a stochastic microswimmer (without any prescribed deformation) can also acquire 
net locomotion due to thermal fluctuations, 
and its average velocity is proportional to the heat flow between the spheres~\cite{Hosaka17}.
Later, the average entropy production rate~\cite{Sou19,Sou21} and the time-correlation functions~\cite{Li24} 
of the same model were calculated by some of the current authors. 
In particular, the existence of an antisymmetric part of the cross-correlation function reflects the 
broken time-reversal symmetry of this microswimmer~\cite{Li24}.

Prior to the above thermal three-sphere microswimmer model, Kumar \textit{et al.}\ proposed 
a model of an active elastic dimer (AED) in which the friction coefficients of the two spheres depend on their 
relative coordinate~\cite{Kumar08,Baule08}. 
Their model is different from the traditional Brownian ratchet models~\cite{Julicher97,Reimann} because the 
motion asymmetry is created internally and is not induced by an external periodic potential.
They found that the average velocity of an AED is proportional to the difference between 
the non-equilibrium noise strengths acting on the two spheres~\cite{Kumar08,Baule08}. 
Such a self-propulsion mechanism can explain, for example, the movement of helicases on 
DNA~\cite{Yu06}, the walking of Myosin VI on actin filaments~\cite{Altman04}, and 
the collective migration of cell clusters~\cite{Pages22}.

Given the modern framework of stochastic thermodynamics~\cite{SekimotoBook,Jarzynski11,Seifert12} and 
the accumulated knowledge of the thermal three-sphere microswimmer model~\cite{Hosaka17,Sou19,Sou21,Li24}, 
the AED model offers a renewed interest, especially when the two spheres have different temperatures 
without any non-equilibrium noises.  
Moreover, the internal feedback control through the position-dependent friction coefficients in the AED model 
is a crucial mechanism for ``informational active matter" that utilizes information (measurement and feedback) 
instead of energy for various non-equilibrium processes~\cite{VanSaders24,Huang20}. 
Recently, we have proposed models of Ornstein-Uhlenbeck information swimmers with external and internal 
feedback controls~\cite{Hou24}.

In this work, we simplify the original AED model to discuss a thermally driven two-sphere microswimmer 
with internal feedback control that is realized by the position-dependent friction coefficients.
In the current model, the two spheres are in thermal equilibrium with independent heat baths having 
different temperatures.
We show that a combination of heat transfer between the spheres and internal feedback control leads to 
directional locomotion in a steady state under a noisy environment.
We analytically obtain the average velocity and the entropy production rate of this stochastic micromachine
by using stochastic energetics~\cite{SekimotoBook}. 
Using these results, we further estimate the efficiency of the proposed model.
We consider that our model explains one of the fundamental mechanisms for transport and locomotion in 
various active matter and biological systems, such as the crawling motion of a cell~\cite{Leoni17,Tarama18}.

In Sec.~\ref{model}, we explain the model of a thermally driven elastic two-sphere microswimmer with internal 
feedback control.
In Sec.~\ref{distribution}, we obtain the steady-state probability distribution that is further used to calculate 
the average velocity and the entropy production rate in Secs.~\ref{velocity} and \ref{epr}, respectively.
In Sec.~\ref{example}, we estimate the efficiency of the two-sphere microswimmer by using specific functional 
forms for the spring potential and the friction coefficients.    
A brief summary and discussion are given in Sec.~\ref{summary}.

\begin{figure}[tb]
\centering
\includegraphics[scale=0.58]{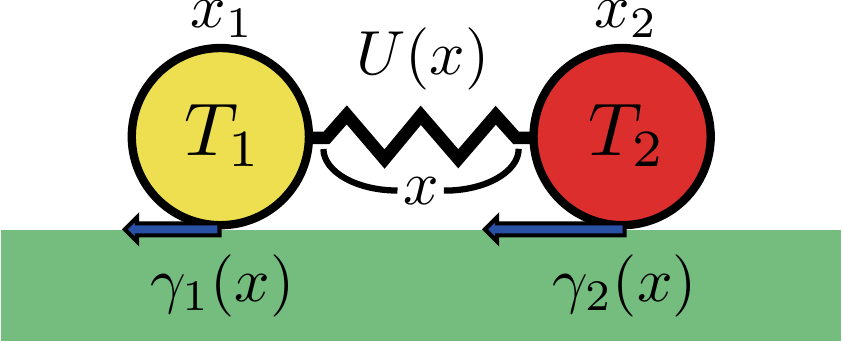}
\caption{(Color online) Thermally driven elastic microswimmer consisting of two hard spheres connected by a spring.
The positions of the two spheres in a one-dimensional coordinate are denoted by $x_i(t)$ ($i=1,2$),  
and the spring extension is defined by $x=x_2-x_1$.
The spring potential energy is given by $U(x)$, and the two spheres are in thermal equilibrium with 
independent heat baths having different temperatures $T_i$.
The friction coefficients $\gamma_i (x)$, acting separately on the two spheres, are 
dependent on the spring extension $x$.
The internal heat flow combined with the feedback control through the position-dependent friction coefficient 
leads to persistent locomotion of the microswimmer.
}
\label{fig_model}
\end{figure}

\section{Model}
\label{model}

As schematically shown in Fig.~\ref{fig_model}, we consider a microswimmer consisting of two 
hard spheres connected by a spring. 
The positions of the two spheres in a one-dimensional coordinate are denoted by $x_i(t)$ ($i=1,2$), 
and we further define the spring extension by $x=x_2-x_1$ that can take both positive and negative values.
The spring potential energy is generally given by $U(x)$, which can be a harmonic potential (as 
we will discuss later) or that of a finitely extensible non-linear elastic (FENE) spring~\cite{Baule08}. 
There are two important assumptions of the model: (i) the two spheres are in equilibrium with 
independent heat baths having different temperatures $T_i$ ($T_1\neq T_2$) and 
(ii) the friction coefficients $\gamma_i (x) \ge 0$, acting separately on the two spheres, are 
dependent on the spring extension $x$.
The fact that the friction coefficients $\gamma_i(x)$ depend on the spring extension $x$ reflects an 
internal feedback mechanism: as the relative position of the spheres changes, the resistance experienced 
by each sphere varies accordingly. 
This coupling allows the system to modulate its dissipative response based on internal configuration 
and is essential for achieving directional locomotion in the absence of external forcing.
To preserve the generality of the model, we first do not specify the functional forms 
of $U(x)$ and $\gamma_i(x)$.

The coupled overdamped Langevin equations for $x_i$ ($i=1,2$) under thermal noise are given by 
\begin{align}
\gamma_i (x) \dot{x}_i(t) = -\partial_i U(x)+g_i(x) \xi_i(t),
\label{GULEq}
\end{align}
where $\dot{x}_i = dx_i/dt$, $\partial_i=\partial/\partial x_i$ 
(the summation over $i$ is not taken here).
The thermal noise $\xi_{i}(t)$ has 
Gaussian statistics with zero mean and unit variance:
\begin{align}
\langle\xi_{i}(t)\rangle=0, \quad  
\langle\xi_{i}(t) \xi_{j}(t')\rangle=\delta_{ij}\delta(t-t').
\label{noise}
\end{align}
According to the fluctuation-dissipation relation of the second kind, the prefactor of thermal noise is 
given by $g_i(x)=\sqrt{2\gamma_i(x)k_{\rm B}T_i}$, where $k_{\rm B}$ is the Boltzmann 
constant~\cite{DoiBook}.
The above Langevin equations are multiplicative because the noise is coupled non-linearly with 
the stochastic variables~\cite{Kampenbook}.
Then, it is necessary to specify the interpretation of the multiplicative noise (otherwise, Eq.~(\ref{GULEq}) 
is meaningless).

Here, we employ the It\^{o} interpretation~\cite{Gardinerbook}, and the Langevin equation 
in Eq.~(\ref{GULEq}) should be rewritten with the It\^{o} convention $x^\ast = x(t)$ as~\cite{Baule08} 
\begin{align}
\dot{x}_i(t)=-\dfrac{\partial_iU(x)}{\gamma_i(x)}
-\dfrac{ [g_i(x)]^2 \partial_i\gamma_i(x)}{2 [\gamma_i(x)]^3}
+\dfrac{g_i(x^\ast)}{\gamma_i(x^\ast)}\xi_i(t).
\label{GOLEq}
\end{align}
As pointed out by Lau and Lubensky~\cite{Lau07}, the second term on the right-hand side 
is proportional to the temperature $T_i$, and it is necessary to ensure the proper thermal 
equilibrium.
In principle, one can also choose other interpretations of the multiplicative noise, such as the Stratonovich 
interpretation~\cite{Gardinerbook}, as long as the same thermal equilibrium is ensured. 
Using, for example, the Stratonovich convention $x^{\ast\ast} = [x(t+\Delta t) + x(t)]/2$, where $\Delta t$ 
is the small increment of time, the corresponding Langevin equation should read
(see the Appendix of Ref.~\cite{Baule08}) 
\begin{align}
\dot{x}_i(t)=-\dfrac{\partial_iU(x)}{\gamma_i(x)}
-\dfrac{ g_i(x) \partial_i g_i(x)}{2 [\gamma_i(x)]^2}
+\dfrac{g_i(x^{\ast\ast})}{\gamma_i(x^{\ast\ast})}\xi_i(t).
\label{stratonovich}
\end{align}
Notice that Eqs.~(\ref{GOLEq}) and (\ref{stratonovich}) statistically describe the same physical phenomena
experiencing a multiplicative noise, and the second terms on the right-hand side of these equations vanish when 
the noise is additive, namely, when $\gamma_i(x)$ is constant.  
A more general argument, including the anti-It\^o convention, can be seen in the literatures~\cite{Lau07,Kuroiwa14}.

For the sake of mathematical convenience, we shall use the It\^{o} interpretation and deal with 
the coupled Langevin equations in Eq.~(\ref{GOLEq}) that can be explicitly written as 
\begin{align}
\dot{x}_1(t)&=\dfrac{U'(x)}{\gamma_1(x)}
+\dfrac{k_{\mathrm{B}}T_1\gamma'_1(x)}{\gamma^2_1(x)}
+\sqrt{\dfrac{2k_{\mathrm{B}}T_1}{\gamma_1(x^\ast)}}\xi_1(t),
\label{ODLEq1}\\
\dot{x}_2(t)&=-\dfrac{U'(x)}{\gamma_2(x)}
-\dfrac{k_{\mathrm{B}}T_2\gamma'_2(x)}{\gamma^2_2(x)}
+\sqrt{\dfrac{2k_{\mathrm{B}}T_2}{\gamma_2(x^\ast)}}\xi_2(t).
\label{ODLEq2}
\end{align}
Hereafter, the prime denotes a derivative with respect to the relative coordinate 
$x=x_2-x_1$, such as $U'(x) = - \partial_1U(x)=\partial_2 U(x)$.

\section{Steady-State Probability Distribution}
\label{distribution}

From the Langevin equations in Eqs.~(\ref{ODLEq1}) and (\ref{ODLEq2}), the equations of motion for both the 
center-of-mass coordinate $X=(x_1+x_2)/2$ and the relative coordinate $x=x_2-x_1$ can be obtained.  
Since the former equation is a function of the relative coordinate $x$ and the noise only, 
we first discuss the equation of motion for $x$ within the It\^{o} interpretation, which is given by~\cite{Baule08}
\begin{align}
\dot{x}(t)=a(x)+b(x^\ast) \xi(t). 
\label{GRCLEq1}
\end{align}
Here, $\xi(t)$ is the superposition of $\xi_{1}(t)$ and $\xi_{2}(t)$ with the same Gaussian 
white statistics, and the two functions $a(x)$ and $b(x)$ are defined as
\begin{align}
a(x)&=-\left(\dfrac{1}{\gamma_1(x)}
+\dfrac{1}{\gamma_2(x)}\right) U'(x)
+k_{\mathrm{B}}T_{1}\left(\dfrac{1}{\gamma_1(x)}
+\dfrac{\theta}{\gamma_2(x)}\right)',
\label{AFu_a(x)} \\
b(x)&=\sqrt{2k_{\mathrm{B}}T_1
\left(\dfrac{1}{\gamma_1(x)}
+\dfrac{\theta}{\gamma_2(x)}\right)},
\label{AFu_b(x)}
\end{align}
where we have introduced the dimensionless temperature ratio $\theta=T_{2}/T_{1}$. 
Note that $\theta=1$ corresponds to the thermal equilibrium.

The Fokker-Planck equation for the probability distribution $P(x,t)$ corresponding to the Langevin 
equation in Eq.~(\ref{GRCLEq1}) can be written in the It\^o interpretation as~\cite{Riskenbook} 
\begin{align}
\dfrac{\partial}{\partial t} P(x,t)
=-\dfrac{\partial}{\partial x}\left[
a(x)P(x,t)\right]+\dfrac{1}{2}
\dfrac{\partial^{2}}{\partial x^{2}}
\left[ b^{2}(x)P(x,t)\right ].
\label{GFPEq}
\end{align}
Since we are interested in the steady-state properties of the microswimmer, we discuss 
the steady-state probability distribution by imposing $\partial_t P(x,t)=0$.
Moreover, the probability flux can be set to zero since the relative coordinate is bounded 
by the spring potential $U(x)$.
Then, the steady-state probability distribution $P_{\rm s}(x)$ satisfies the relation~\cite{Baule08} 
\begin{align}
a(x)P_{\mathrm{s}}(x)-\dfrac{1}{2}\dfrac{\partial}{\partial x}
\left[b^{2}(x)P_{\mathrm{s}}(x)\right]=0,
\label{SFPEq}
\end{align}
and $P_{\rm s}(x)$ can be formally solved as  
\begin{align}
P_{\mathrm{s}}(x)=\dfrac{\mathcal{N}}{b^{2}(x)}
\exp\left[ \int_{0}^{x}dy \, \dfrac{2a(y)}{b^{2}(y)} \right],
\label{GSSPD1}
\end{align}
where $\mathcal{N}$ is the normalization factor. 
By substituting Eqs.~(\ref{AFu_a(x)}) and (\ref{AFu_b(x)}) into the above expression, we obtain~\cite{Baule08} 
\begin{align}
P_{\mathrm{s}}(x)=\dfrac{\mathcal{N}_1}{2k_{\mathrm{B}}T_1}\exp\left[
\int_{0}^{x} dy \, \left(-\dfrac{U'(y)[\gamma^{-1}_1(y)+\gamma^{-1}_2(y)]}
{k_{\mathrm{B}}T_1[\gamma^{-1}_1(y)+\theta\gamma^{-1}_2(y)]}
\right)\right].
\label{GSSPD2}
\end{align}
If we further assume that the two friction coefficients are identical for all $x$, i.e., 
$\gamma_1(x)=\gamma_2(x)$, the steady-state probability distribution simplifies to 
\begin{align}
P_{\mathrm{s}}(x)=\dfrac{\mathcal{N}_1}{2k_{\mathrm{B}}T_1}
\exp\left[ -\dfrac{2U(x)}{(1+\theta)k_{\mathrm{B}}T_1} \right],
\label{SSPD1}
\end{align}
which is essentially the Boltzmann distribution modified by the temperature ratio $\theta=T_2/T_1$.

\section{Average Velocity}
\label{velocity}

Having obtained the steady-state distribution function $P_{\mathrm{s}}(x)$ for the relative coordinate $x$
in the previous section, we calculate the average velocity of the elastic two-sphere microswimmer.
Since the center of mass velocity is simply given by $V=\dot{X}=(\dot{x}_{1}+\dot{x}_{2})/2$, we use 
Eqs.~(\ref{ODLEq1}) and (\ref{ODLEq2}) to obtain the statistical average of $V$~\cite{Kumar08,Baule08} 
\begin{equation}
\langle V\rangle
=\frac{1}{2}\left\langle \left(\dfrac{1}{\gamma_1(x)}
-\frac{1}{\gamma_2(x)}\right) U'(x) \right\rangle
-\frac{k_{\mathrm{B}}T_1}{2}
\left\langle \left(\dfrac{1}{\gamma_{1}(x)}
-\frac{\theta}{\gamma_{2}(x)} \right)' \right\rangle,
\label{GAVEq1}
\end{equation}
where the averaging $\langle \cdots\rangle$ is performed over the steady-state distribution $P_{\rm s}(x)$.
To derive the above expression, the contribution of the multiplicative noise terms has been omitted owing 
to the It\^o interpretation.
It should be emphasized here that $\langle V\rangle$ vanishes when $\theta=1$ for any 
choice of $\gamma_1(x)$ and $\gamma_2(x)$, provided that the average is evaluated with
respect to $P_{\mathrm{s}}(x)$.

When $\gamma_1(x)=\gamma_2(x)$, the above average velocity further reduces to 
\begin{align}
\langle V \rangle
=\dfrac{(\theta-1)k_{\mathrm{B}}T_{1}}{2}
\left\langle\left(\dfrac{1}{\gamma_1(x)}\right)' \right\rangle.
\label{AVEq1}
\end{align}
This result clearly demonstrates that the microswimmer acquires a finite average velocity 
when $\theta \neq 1$ ($T_1 \neq T_2$) and $\gamma_1$ is not constant.  
In other words, the current microswimmer is driven by thermal energy, and the internal feedback 
control through $\gamma_1(x)$ plays a crucial role for its locomotion.
We remark, however, that when $(1/\gamma_1(x))'$ and $U(x)$ are respectively odd and even functions of $x$, 
$\langle V \rangle$ vanishes even if the friction coefficient is position-dependent.  
A similar result to Eqs.~(\ref{GAVEq1}) and (\ref{AVEq1}) was reported for an AED whose asymmetry in the 
non-equilibrium noise strengths leads to locomotion~\cite{Kumar08,Baule08}. 
In our study, we do not introduce any active fluctuations, which allows us to use the standard framework 
of stochastic thermodynamics~\cite{SekimotoBook,Jarzynski11,Seifert12}, as we discuss below.

\begin{figure}[tb]
\centering
\includegraphics[scale=0.75]{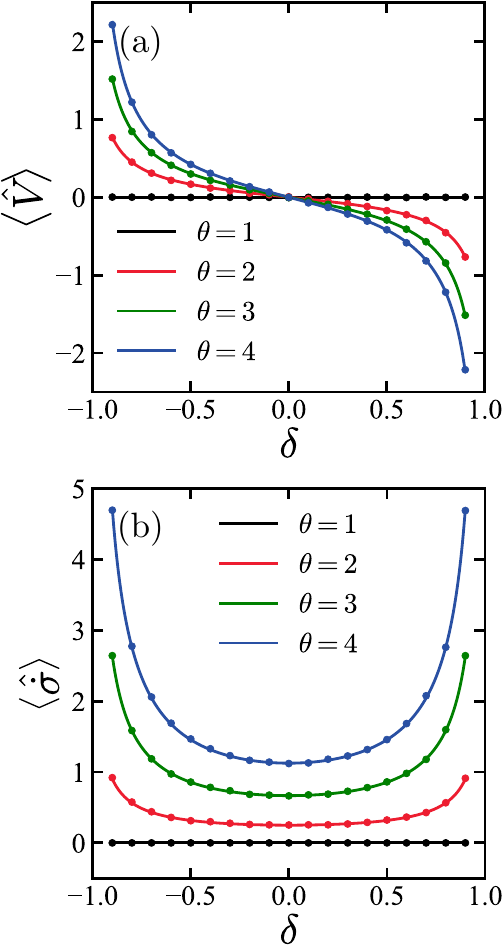}
\caption{(Color online) The plots of the average velocity $\langle V \rangle$ [see Eq.~(\ref{SAVEq1})] and 
the average entropy production rate $\langle \dot{\sigma}\rangle$ [see Eq.~(\ref{SHFEq_1})]
of the two-sphere microswimmer when the potential is $U(x)=Kx^2/2$ and the 
friction coefficients are $\gamma_1(x)=\gamma_2(x)=\gamma[1+\delta\tanh(x/w)]$ 
[see Eq.~(\ref{frictionexample})].
The dimensionless energy parameter is chosen here as $\varepsilon= Kw^2/(k_{\mathrm{B}}T_1)=1$.
(a) The plot of the dimensionless average velocity 
$\langle \hat V \rangle = \langle V \rangle \gamma /\sqrt{K k_{\mathrm{B}}T_1}$
as a function of the dimensionless feedback strength parameter $\delta$ for different values 
of the dimensionless temperature ratio $\theta=T_2/T_1$. 
(b) The plot of the dimensionless average entropy production rate 
$\langle\hat{\dot{\sigma}}\rangle=\langle \dot{\sigma}\rangle \gamma T_1/(Kw)^2$ 
as a function of $\delta$ for different values of $\theta$.
In both (a) and (b), the filled circles correspond to the result of the numerical simulation. 
}
\label{fig_velocity}
\end{figure}

\begin{figure}[tb]
\centering
\includegraphics[scale=0.5]{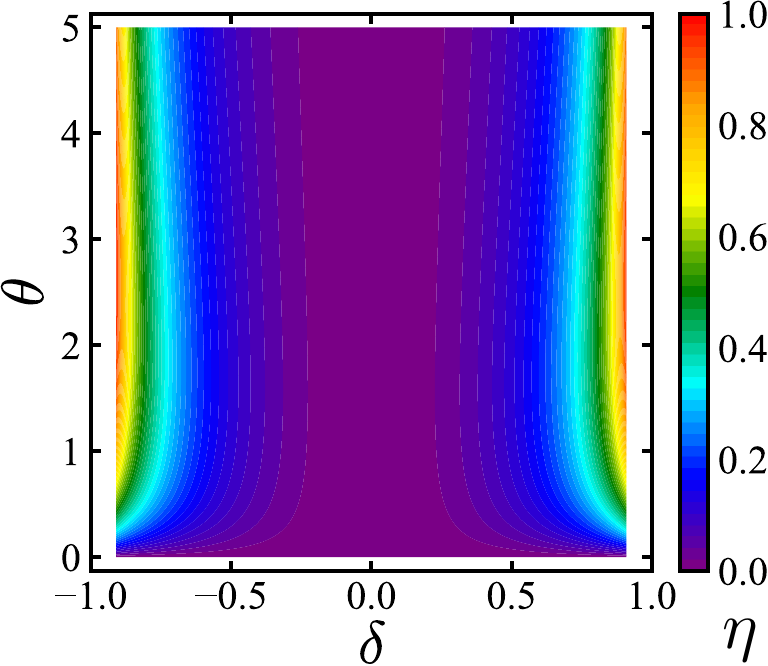}
\caption{(Color online) The color plot of the efficiency $\eta$ [see Eq.~(\ref{efficiency})] of the two-sphere 
microswimmer as a function of the dimensionless feedback strength parameter $\delta$ and 
the dimensionless temperature ratio $\theta=T_2/T_1$.
The efficiency $\eta$ vanishes when $\delta=0$. 
}
\label{fig:efficiency}
\end{figure}

\section{Entropy Production Rate}
\label{epr}

In this section, we calculate the average entropy production rate of the elastic microswimmer in the steady state, 
which is given by
\begin{align}
\langle\dot{\sigma} \rangle
=-\frac{\langle\dot{Q}_1 \rangle}{T_1} - \dfrac{\langle\dot{Q}_2 \rangle}{T_2},
\label{GEPREq1}
\end{align}
where the stochastic heat flow $\dot{Q}_i$ is the heat gained by the $i$-th sphere per unit time. 
According to the stochastic energetics developed by Sekimoto~\cite{SekimotoBook}, it is given by
\begin{align}
\dot{Q}_i \, dt =\dfrac{\partial U(x)}{\partial x_{i}}\circ dx_i,
\label{GHEq}
\end{align}
where $\circ$ denotes the Stratonovich product, and the small increment $dx_i$ can be obtained 
from the Langevin equations in Eqs.~(\ref{ODLEq1}) and (\ref{ODLEq2}).
Using the Wong-Zakai theorem to deal with the Stratonovich product and taking the 
statistical average~\cite{SekimotoBook}, we obtain after some calculation 
\begin{align}
\langle \dot{Q}_i \rangle
&=-\left\langle\dfrac{[U'(x)]^2}{\gamma_i(x)}\right\rangle
+k_{\mathrm{B}}T_i\left\langle \left(\dfrac{U'(x)}{\gamma_i(x)}\right) '\right\rangle.
\label{GHFEq_1}
\end{align}
A similar expression was obtained in Eq.~(32) of Ref.~\cite{Cates22}.
When the two friction coefficients are constant and identical, the above expression reduces to 
Eq.~(4.27) of Ref.~\cite{SekimotoBook}.

It is straightforward to verify that $\langle \dot{Q}_i \rangle = 0$ when $\theta =1$ by using 
the equilibrium distribution function in Eq.~(\ref{GSSPD2}). 
When $\theta \neq1$, we now consider the total heat flow defined by 
$\langle\dot{Q}\rangle =\langle\dot{Q}_1\rangle+\langle\dot{Q}_2\rangle$.
Using Eqs.~(\ref{AFu_a(x)}) and (\ref{AFu_b(x)}), one can generally show that 
$\langle\dot{Q}\rangle$ vanishes in the steady state because 
\begin{align}
\langle\dot{Q}\rangle=
\langle U'(x)a(x)\rangle +\dfrac{1}{2}\langle U''(x)b^{2}(x)\rangle=0.
\label{THFEq}
\end{align}
In the second equality, we have used Eq.~(\ref{SFPEq}), which stems from the steady-state condition. 
Notice that the relation $\langle\dot{Q}\rangle=0$ generally holds even if the two friction coefficients 
$\gamma_1(x)$ and $\gamma_2(x)$ are different. 
Since we have $\langle\dot{Q}_2\rangle= -\langle\dot{Q}_1\rangle$ from Eq.~(\ref{THFEq}), 
the average entropy production rate in Eq.~(\ref{GEPREq1}) can be rewritten as 
\begin{align}
\langle\dot{\sigma}\rangle=\frac{1-\theta}{\theta} \frac{\langle\dot{Q}_{1}\rangle}{T_1},
\label{EPREq1}
\end{align}
which vanishes when $\theta=1$. 
When $T_1 < T_2$ ($\theta >1$), $\langle\dot{Q}_1\rangle$ is negative and we have 
$\langle\dot{\sigma}\rangle >0$, as required from the generalized second law for 
non-equilibrium systems. 
A similar statement also holds when $T_1 > T_2$ ($\theta <1$).

\section{Example}
\label{example}

So far, we have not yet specified the functional forms of the spring potential $U(x)$ and the friction 
coefficients $\gamma_i(x)$.
As an example, we choose the harmonic potential energy $U(x)=Kx^2/2$, where $K>0$ is the
spring constant, and assume that the friction coefficients have the form~\cite{Kumar08,Baule08} 
\begin{align}
\gamma_1(x)=\gamma_2(x)=\gamma[1+\delta\tanh(x/w)], 
\label{frictionexample}
\end{align}
where $\gamma$ is a constant
friction coefficient, $\delta$ is a dimensionless feedback strength parameter satisfying $\vert \delta \vert < 1$ (recall 
$\gamma_i(x) > 0$), and $w$ is a length characterizing the change of the friction 
coefficients. 
In this case, the steady-state distribution function in Eq.~(\ref{SSPD1}) and the average velocity
in Eq.~(\ref{AVEq1}) become 
\begin{align}
P_{\mathrm{s}}(x)=\sqrt{\dfrac{K}
{\pi(1+\theta)k_{\mathrm{B}}T_1}}
\exp \left[-\dfrac{Kx^2}{(1+\theta)k_{\mathrm{B}}T_1} \right],
\label{SSSPD}
\end{align}
and 
\begin{align}
\langle V \rangle
& =\dfrac{\delta(1-\theta)}{2\gamma}
\sqrt{\dfrac{Kk_{\mathrm{B}}T_1}{\pi(1+\theta)}}
\nonumber\\
&\times 
\int_{-\infty}^{\infty} dz \,
\dfrac{\sech^2 z}
{\left(1+\delta\tanh z\right)^2} 
\exp \left( -\dfrac{\varepsilon z^2}{1+\theta}  \right),
\label{SAVEq1}
\end{align}
respectively, where $\varepsilon = Kw^2/(k_{\mathrm{B}}T_1)$ is the dimensionless energy parameter.
Note that Eq.~(\ref{SSSPD}) is simply a Gaussian distribution function. 
Since the dimensionless integral in Eq.~(\ref{SAVEq1}) is positive (which will be evaluated numerically below), 
$\langle V \rangle$ is non-zero only when $\delta \neq 0$ and $\theta \neq 1$. 
Moreover, the temperature ratio $\theta$ determines the direction of locomotion, i.e., 
$\langle V \rangle <0$ when $\delta >0$ and $\theta >1$, or vice versa. 
Similarly, by calculating the average heat flow $\langle\dot{Q}_{1}\rangle$ in Eq.~(\ref{GHFEq_1}), 
we obtain the average entropy production rate in Eq.~(\ref{EPREq1}) as 
\begin{align}
\langle \dot{\sigma} \rangle
& =  \frac{\theta-1}{\theta} \frac{(Kw)^2}{\gamma T_1} \sqrt{\dfrac{\varepsilon}{\pi(1+\theta)}} 
\int_{-\infty}^{\infty} dz \,
\left[ \dfrac{z^2}{1+\delta\tanh z} 
\right.
\nonumber \\ 
& \left.
+\dfrac{\delta z\sech^2 z}{\varepsilon(1+\delta\tanh z)^2}
-\dfrac{1}{\varepsilon (1+\delta\tanh z)}\right]
\exp\left(-\dfrac{\varepsilon z^2}{1+\theta} \right),
\label{SHFEq_1}
\end{align}
which includes another dimensionless integral.

In Fig.~\ref{fig_velocity}(a), we numerically plot the dimensionless average velocity 
$\langle \hat V \rangle = \langle V \rangle \gamma /\sqrt{K k_{\mathrm{B}}T_1}$ from Eq.~(\ref{SAVEq1})
as a function of the feedback strength parameter $\delta$ for different values of the temperature ratio 
$\theta=T_2/T_1 >1$ when $\varepsilon=1$. 
To confirm our analytical prediction, we have also performed numerical simulations of the coupled stochastic 
equations in Eqs.~(\ref{ODLEq1}) and (\ref{ODLEq2}). 
These Langevin equations were discretized according to the It\^o convention.
We find a good agreement between the theoretical and numerical results.
Notice that $\langle \hat V \rangle$ vanishes when either $\theta=1$ or $\delta =0$, as shown in Eq.~(\ref{SAVEq1}).
Moreover, $\langle \hat V \rangle$ is an odd function of $\delta$ because the sign of $\delta$ determines 
the direction of locomotion.
In Fig.~\ref{fig_velocity}(b), on the other hand, we numerically plot the dimensionless average entropy production rate 
$\langle\hat{\dot{\sigma}}\rangle=\langle \dot{\sigma}\rangle \gamma T_1/(Kw)^2$ from Eq.~(\ref{SHFEq_1})  
as a function of the feedback strength $\delta$.
For $\theta > 1$, $\langle \dot{\sigma}\rangle$ is positive and is an even function of $\delta$. 
As a result, the entropy production rate is minimized at $\delta =0$ and the corresponding 
expression is given by 
$\langle\dot{\sigma}\rangle_{\rm min}=k_{\mathrm{B}}K(\theta-1)^2/(2\gamma\theta) \ge 0$.

At this stage, it is worth evaluating the efficiency of the current microswimmer defined 
by~\cite{Golestanian08,Sou21} 
\begin{align}
\eta=\dfrac{2\gamma\langle V\rangle^{2}}{\langle\dot{\sigma}\rangle T^{\ast}},
\label{efficiency}
\end{align}
where $T^{\ast}=(T_1+T_2)/2$ is the average temperature although either $T_1$ or $T_2$ can also be used. 
The factor $2\gamma$ in the numerator reflects the fact that the microswimmer consists of two spheres 
with the same average friction coefficient $\gamma$.  
In Fig.~\ref{fig:efficiency}, we color-plot the efficiency $\eta$ as a function of the feedback 
strength parameter $\delta$ and the temperature ratio $\theta=T_2/T_1$.
The efficiency $\eta$ vanishes when $\delta=0$ and is an even function of $\delta$ when $\theta$ is fixed.
It is worth noting that, in the equilibrium limit of $\theta \rightarrow 1$, the efficiency $\eta$ is finite because 
both $\langle V \rangle$ and $\langle\dot{\sigma}\rangle$ vanish in this limit [see Eqs.~(\ref{SAVEq1}) and (\ref{SHFEq_1})].

In general, there is a contribution to the total entropy production rate coming from the center-of-mass 
diffusion~\cite{Sou21}, which is neglected in Eq.~(\ref{efficiency}).
Here, we have only considered the entropy production rate due to the heat flow [see Eq.~(\ref{GEPREq1})]
that causes the unidirectional locomotion.

\section{Summary and Discussion}
\label{summary}

To summarize, we have studied the locomotion of a thermally driven elastic two-sphere microswimmer that has 
position-dependent friction coefficients. 
Such a microswimmer can acquire non-zero average velocity in the steady state due to the heat flow between 
the spheres.
We have obtained the entropy production rate and further estimated the efficiency of the microswimmer. 
The proposed self-propulsion mechanism emphasizes the importance of active matter and biological systems 
driven by internal feedback control.
In the future, we will calculate the time-correlation functions of the microswimmer to quantitatively discuss the 
degree of broken time-reversal symmetry~\cite{Li24}.

In our model, the state-dependent friction coefficients play an essential role. 
The change in the friction coefficient can be caused, for example, by the change in the particle size. 
In this sense, the current model has a similarity to the model of ``pushmepullyou" by Avron \textit{et al.}~\cite{Avron05} 
or the two-sphere model by Pandey \textit{et al.}~\cite{Pandey12}.  
The main difference in our model is that the spheres have different temperatures and thermal fluctuations cause 
the locomotion in a stochastic manner. 
The internal heat flow combined with the feedback control through the position-dependent friction coefficient 
is the main driving force for persistent locomotion.

The other important assumption of the model is that the statistical properties of the two spheres are characterized by 
two different temperatures.
Such a micromachine is conceptually motivated by Feynman's Brownian ratchet~\cite{Feynmanbook} or a model system that 
interacts with two thermal environments~\cite{SekimotoBook}.
More recently, a similar non-equilibrium dimer model confined between two walls was discussed to calculate probability flux loops 
that demonstrate the broken detailed balance when the temperatures are different~\cite{Battle16,Gnesotto18,Li19}.
In the previous thermally driven three-sphere microswimmer model~\cite{Hosaka17}, all the spheres were assumed to have 
different temperatures.

Although it is technically challenging to introduce different temperatures in a moving microswimmer, one possible way is to
use a chemically heterogeneous particle, such as a Janus particle, under laser irradiation, which induces inhomogeneous 
temperature distribution on the particle~\cite{Li19}. 
By further introducing the internal deformation degree of freedom~\cite{Yang14}, we expect that the proposed thermally 
driven two-sphere microswimmer can be experimentally realized. 
It should be mentioned, however, that any difference in the fluctuation (both thermal and non-thermal) between the two 
spheres is sufficient to induce a non-zero velocity of the two-sphere microswimmer~\cite{Kumar08,Baule08}.
Hence, rather than the thermodynamic temperature itself, an effective temperature that reflects the microswimmer's internal 
fluctuation is more important for its persistent locomotion.

\begin{acknowledgments}

We thank Z.\ Xiong for the useful discussions.
Z.H., L.H.\, and S.K.\ acknowledge the support by the National Natural Science Foundation of China 
(Nos.\ 12104453, 22273067, 12274098, and 12250710127).
Y.H.\ acknowledges support from JSPS Overseas Research Fellowships (Grant No.  202460086).
S.K.\ acknowledges the startup grant of Wenzhou Institute, 
University of Chinese Academy of Sciences (No.\ WIUCASQD2021041). 
K.Y\ and S.K.\ acknowledge the support by the Japan Society for the Promotion of Science (JSPS) Core-to-Core 
Program ``Advanced core-to-core network for the physics of self-organizing active matter" (No.\ JPJSCCA20230002).
J.L.\ and Z.Z.\ contributed equally to this work.
\end{acknowledgments}


\end{document}